\documentclass[prb,twocolumn,showpacs,amsmath,amssymb]{revtex4}

\usepackage{graphicx}
\usepackage{dcolumn}
\usepackage{bm}
\usepackage{epsfig}

\begin{document}

\title{Quantum Transparency of Anderson Insulator Junctions: \\ Statistics of Transmission Eigenvalues, Shot Noise, and Proximity Conductance}

\author{Branislav K. Nikoli\' c and Ralitsa L. Dragomirova}
\affiliation{Department of Physics and Astronomy, University
of Delaware, Newark, DE 19716-2570}

\begin{abstract}
We investigate quantum transport through strongly disordered barriers, made of a material with exceptionally high resistivity that behaves as an Anderson insulator or a ``bad metal'' in the bulk, by analyzing the distribution of Landauer transmission eigenvalues  for a junction where such barrier is attached to two clean metallic leads. We find that scaling of the transmission eigenvalue distribution with the junction thickness (starting from the single interface limit) always predicts a non-zero probability to find high transmission channels even in relatively thick barriers. Using this distribution, we compute the zero frequency shot noise power
(as well as its sample-to-sample  fluctuations) and demonstrate how it provides
a single number characterization of non-trivial transmission properties of
different types of disordered barriers. The appearance of open conducting
channels, whose transmission eigenvalue is close to one, and corresponding
violent mesoscopic fluctuations of transport quantities explain at least some
of the peculiar zero-bias anomalies in the Anderson-insulator/superconductor
junctions observed in recent experiments [Phys. Rev. B {\bf 61}, 13037 (2000)].
Our findings are also relevant for the understanding of the role of defects that
can undermine quality of thin tunnel barriers made of conventional band-insulators.
\end{abstract}

\maketitle

\section{Introduction} \label{sec:intro}

The advent of mesoscopic quantum physics~\cite{meso,datta_book,mello} in the early 1980s has profoundly influenced our understanding of transport in solids.  Advances in micro- and nano-fabrication technology have brought about small enough structures ($\lesssim  1\mu$m) in which, at low enough temperatures ($T \ll 1$K), propagation of an electron is described by a single wave function (instead of density matrices in macroscopic solids) since inelastic dephasing processes can be suppressed below the temperature-dependent dephasing length $L_\phi$. Thus, their transport properties have to be analyzed in terms of quantities that take into account  non-local features of quantum dynamics (such as the quantum corrections~\cite{mello} to the conductivity which are non-local on the scale of $L_\phi$), finite-size of the sample, boundaries,
and measurement set-up of macroscopic external circuit, rather than using traditional local and self-averaging quantities (such as the conductivity) applicable to bulk materials at high enough  temperatures.

Particularly influential ideas have emanated from the Landauer-B\" uttiker approach~\cite{mello,datta_book,carlo_rmt} to quantum transport which treats conduction within the phase-coherent sample as a complicated (multichannel) quantum-mechanical scattering problem. This viewpoint introduces a set of transmission coefficients as the fundamental property of a mesoscopic conductor. The transmission coefficients $T_n$ are formally defined as the eigenvalues of ${\bf t} {\bf t}^\dag$, which is the product of a transmission matrix ${\bf t}$ and its Hermitian conjugate ${\bf t}^\dag$. In the two-probe geometry, where mesoscopic sample is attached to two semi-infinite ideal metallic leads, the ${\bf t}$-matrix connects the transmission amplitudes of the flux-normalized states in the left lead to the outgoing states in the right lead.  Thus, the basis of eigenchannels, which diagonalizes the matrix ${\bf t} {\bf t}^\dag$, offers a simple intuitive picture where conductor can be viewed as a parallel circuit of independent transmission channels characterized by channel-dependent transmission probability $T_n$. Within this framework, pure tunnel barrier is a rather simple case where all transmission eigenvalues $T_n \ll 1$ are the same and much smaller than one (the opposite limit, $T_n = 1$, is a property of the ballistic transport occurring through fully open conducting channels).

Since many electronic devices employ quantum-mechanical tunneling through an insulating
barrier, their design and optimization requires to understand whether transport
occurs via pure tunneling or if it is affected also by the defects in the barrier.~\cite{rough,mtjballistic} In particular, high-critical current density
for Josephson tunnel junctions~\cite{jj} or impedance level for magnetic tunnel junctions~\cite{mtj} require ultrathin and highly transparent barriers that can easily be pushed out of the genuine tunneling regime.~\cite{rough} The diagnostics of non-trivial barrier properties requires to investigate quantities beyond just the conductance since its exponential
decrease with the barrier thickness, as a naive criterion of pure tunneling, can be
generated by vastly different underlying microscopic mechanisms. For example, recent experiments~\cite{naveh,rough} have pointed out how homogeneous ultrathin (e.g.,
thickness $\sim 1$nm) aluminum oxide barriers  can accommodate high transmission
channels $T_n \simeq 1$ (which are detrimental for various device operation~\cite{mtjballistic}). This is due to extended states induced by disorder~\cite{rough} or intrinsic transport mechanism in disordered
mesoscopic systems,~\cite{naveh} rather than due to rare defects such as pinholes
with more than unit-cell dimension.

When static disorder becomes strong enough, solids undergo localization-delocalization (LD) transition leading to an Anderson insulator.~\cite{anderson} Such phase is substantially different from the conventional Bloch-Wilson band insulator since density of states at the Fermi energy remains finite in Anderson insulators. On the other hand, the wave function associated with the localized states is confined within the region of a characteristic size specified by the localization length  $\xi$.

Here we explore quantum transport through a strongly disordered barrier, separating the two clean metallic electrodes, by computing statistical properties of the transmission eigenvalues for an ensemble of  three-dimensional (3D) samples with different impurity configuration. We focus on the appearance of completely open transmission channels~\cite{zvi,asano} $T_n \simeq 1$, as the barrier thickness increases from the single interface limit to the junction thickness where tunneling through the Anderson insulator takes place, and their effect on experimentally accessible transport properties.  That is, the full statistics of $T_n$ allows us to obtain frequently measured quantities that contain the signatures of such non-trivial transparency properties: (a) the zero-frequency power spectrum of the shot noise;~\cite{shot_noise} and (b) the conductance~\cite{carlo_rmt} $G_{NS}$  of a hybrid junction composed of a thin Anderson insulator
attached to a superconductor, whose unusual  properties have been unearthed in recent mesoscopic
transport experiments.~\cite{zvi} Our findings on quantum transmissivity of single interface and thin barriers of a strongly disordered materials are relevant also for the analogous classical coherent scattering problems, such
as the light propagation through thin, but strongly diffusive, medium.~\cite{patra}

The paper is organized as follows. In Sec.~\ref{sec:hamilton} we introduce the Hamiltonian model of the disordered barrier and corresponding real-space Green function technique that allows us to obtain an exact transmission matrix of a specific sample. In Sec.~\ref{sec:trans} we study the scaling of the distribution of $T_n$ as a function of the barrier thickness, where disorder strength  serves
as a parameter whose tuning induces Anderson insulator, as well as a ``bad metal'' regime upon approaching the LD transition from the metallic side. Measurable transport quantities---shot noise and proximity conductance $G_{NS}$---determined by these  distributions are discussed in Sec.~\ref{sec:shot}. In particular, we find the shot noise to be a sensitive single parameter characterization of the transparency of multichannel barriers, as well as of different types of diffusion through dirty metallic barriers. We conclude in Sec.~\ref{sec:conclusion}.
\begin{figure}
\centerline{\psfig{file=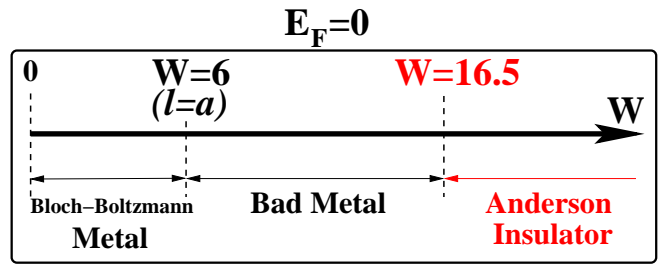,width=3.4in,angle=0} }
\caption{The boundaries of different transport regimes, determined by
the strength of the disorder $W$, in a bulk 3D conductor described by the
half-filled ($E_F=0$) Anderson model. At $W \approx 6$, the Boltzmann equation
breaks down (putative semiclassical mean free path becomes smaller than the lattice
spacing $\ell \le a$), while at $W \approx 16.5$ wave functions become localized.
Within the intermediate ``bad metal'' regime, particle motion is ``intrinsic'' diffusion
that requires non-perturbative quantum description.~\cite{nikolic_rho}} \label{fig:anderson}
\end{figure}

\section{Transmission properties of the Anderson model for disordered barrier} \label{sec:hamilton}

We model non-interacting electrons in the disordered barrier
by a standard Anderson model,~\cite{anderson}
\begin{equation}\label{eq:tbh}
  \hat{H} = \sum_{\bf m} \varepsilon_{\bf m}|{\bf m} \rangle \langle {\bf m}|
  +  t\sum_{\langle {\bf m},{\bf n} \rangle}
  |{\bf m} \rangle \langle {\bf n}|,
\end{equation}
which is a tight-binding Hamiltonian (TBH) defined on a simple cubic lattice
$L \times L_y \times L_z$.  The nearest neighbor hopping matrix element, between
$s$-orbitals $\langle {\bf r}|{\bf m} \rangle = \psi({\bf r}-{\bf m})$ on adjacent
atoms located at sites ${\bf m}$ of the lattice, is denoted by $t$ and sets the
unit of energy. Here $L$ is the thickness of the junction in the units of the lattice spacing $a$ (i.e., $L$ is equal to the number of  disordered interfaces of the cross section $L_y \times L_z$ which are stacked along the $x$-axis, chosen as the direction of transport, and coupled via hopping $t$ to form the barrier). We set $L_y=L_z=20$, which yields the quantum point contact conductance $G_{QPC}(E_F=0)= 259 G_Q$ ($G_Q=2e^2/h$ is the conductance quantum) of the corresponding clean system attached to two leads of the same cross section (i.e., for this set-up, there are at most 259 fully open Landauer conducting channels $T_n=1$ at half-filling, out of 400 supported by its cross section~\cite{nikolic_jcmp}). The disorder is introduced by setting a random on-site potential such that $\varepsilon_{\bf m}$ is uniformly distributed in the interval $[-W/2,W/2]$. The whole band of the Anderson model becomes localized, i.e., the LD transition takes place at the Fermi energy $E_F=0$ of the half-filled band, when critical disorder strength $W_c \approx 16.5$ is reached.

It is important for subsequent discussion to recall that there are three fundamentally different transport
regimes in bulk 3D disordered conductors~\cite{nikolic_rho}  (i.e., in the cubes $L \times L \times L$ with a
given concentration of impurities): (a) the semiclassical regime, where the Bloch-Boltzmann theory and
perturbative quantum corrections (obtained from the Kubo formula) describe resistivity of diffusive ($\ell \ll L$) systems; (b) the ``bad metal'' regime characterized by exceptionally huge resistivities and lack of  semiclassical mean free path $\ell$ (the putative mean free path would be smaller than the lattice spacing $\ell < a$; nevertheless such ``intrinsic'' quantum diffusion can still be described by a diffusion constant extracted from the Kubo formula~\cite{nikolic_rho}), whereby semiclassical description and perturbative methods, based on the expansion in a small parameter $1/k_F\ell$, break down; and (c) the Anderson localized regime when disorder becomes strong enough to push  the conductance of a disordered sample below~\cite{carlo_rmt} $2e^2/h$. Note that to observe the effects stemming from localization of wave functions,
the size of the conductor has to be greater than the localization length $L \gg \xi$---on length scales smaller than $\xi$ one cannot differentiate an Anderson insulator from a disordered metal. Figure~\ref{fig:anderson} delineates the boundaries of these regimes for a system modeled by the
half-filled Anderson Hamiltonian of Eq.~(\ref{eq:tbh}).

The transmission matrix ${\bf t}$
\begin{equation}\label{eq:transmission}
  {\bf t}  =  2 \sqrt{-\text{Im} \, \hat{\Sigma}_L} \cdot \hat{G}^{r}_{1 N}
  \cdot \sqrt{-\text{Im} \, \hat{\Sigma}_R},
\end{equation}
is obtained from the real-space Green function $\hat{G}^{r,a}$
\begin{equation} \label{eq:green}
\hat{G}^{r,a} = \frac{1}{E-\hat{H}-\hat{\Sigma}^{r,a}},
\end{equation}
where $\hat{G}^{r}_{1 N}$, $\hat{G}^{a}_{N 1}$ are submatrices
of $\hat{G}^{r,a}$ ($\hat{G}^{a}=[\hat{G}^{r}]^{\dagger}$)
that connect layers $L=1$ and $L=N$ of the sample along
the $x$-axis. Here
$\text{Im} \, \hat{\Sigma}_{L,R}=(\hat{\Sigma}_{L,R}^r-\hat{\Sigma}_{L,R}^a)/2i$
are self-energy matrices ($r$-retarded, $a$-advanced) which
describe the coupling of the sample to the leads,~\cite{datta_book} with
$\hat{\Sigma}^{r}=\hat{\Sigma}_L^{r}+\hat{\Sigma}_R^{r}$
($\hat{\Sigma}^{a}=[\hat{\Sigma}^{r}]^{\dagger})$. This particular
computationally efficient implementation of the Landauer-B\" uttker
formalism, which takes the microscopic Hamiltonian as an input,  has its
origins in the treatment of tunneling current in metal/insulator/metal ($MIM$)
junctions---it was developed in order to evade pathological properties of
a tunneling Hamiltonian when attempting to take into account higher
order tunneling processes.~\cite{caroli}
\begin{figure}
\centerline{\psfig{file=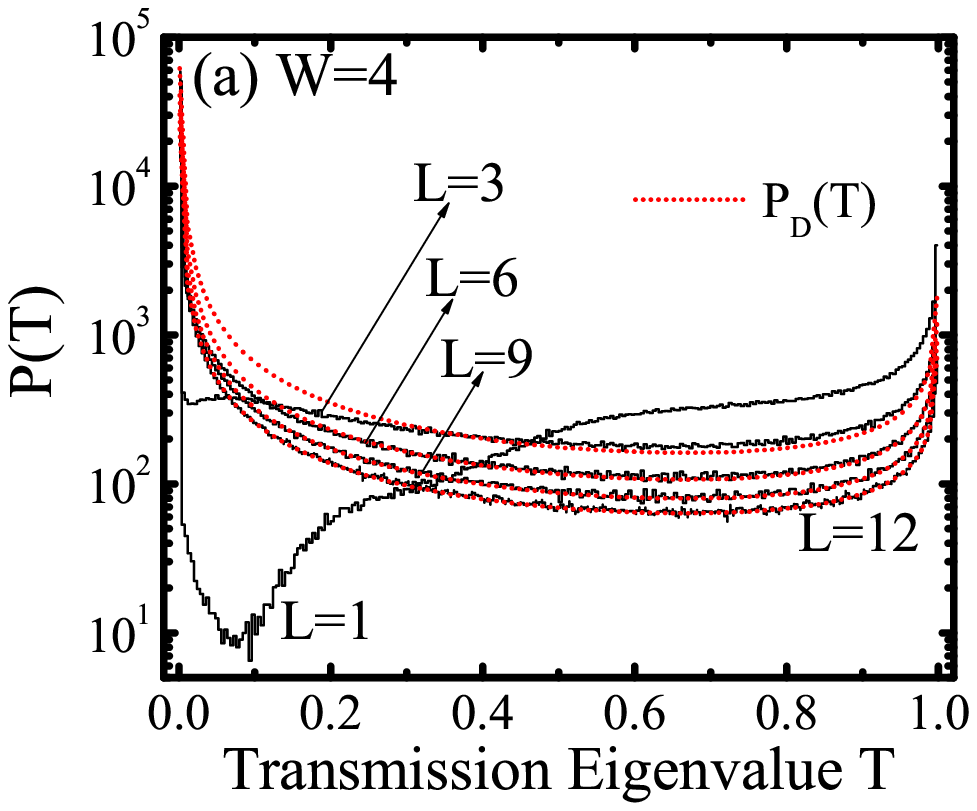,width=3.3in,height=2.5in,angle=0} }
\vspace{-0.1in}
\centerline{\psfig{file=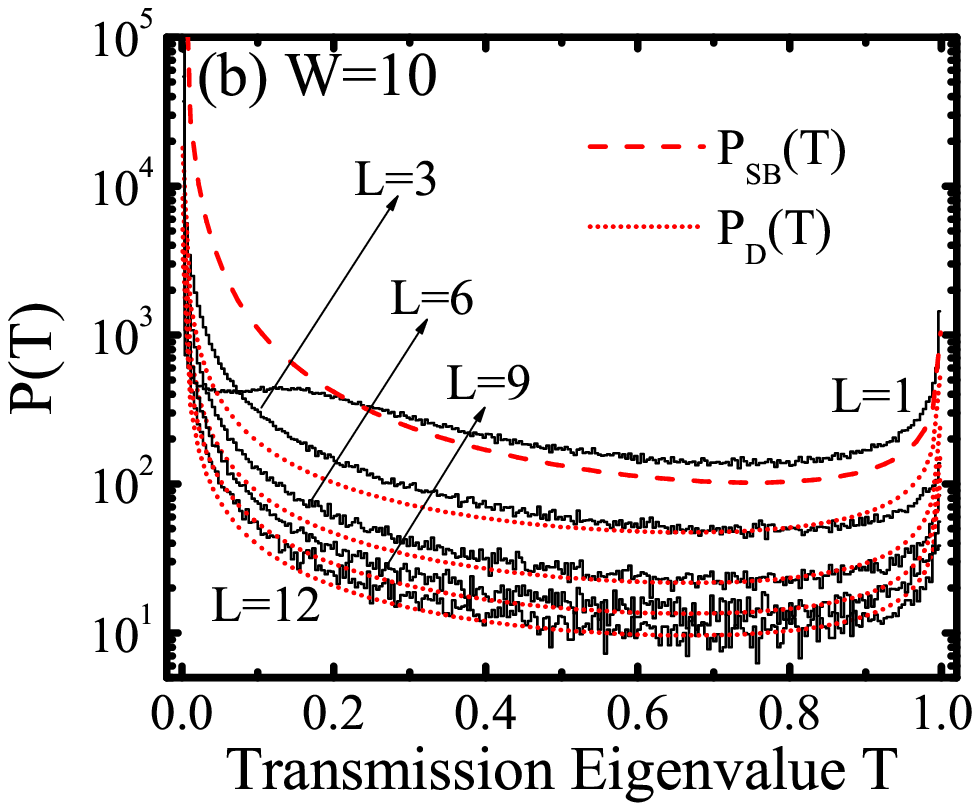,width=3.3in,height=2.5in,angle=0} }
\vspace{-0.1in}
\centerline{\psfig{file=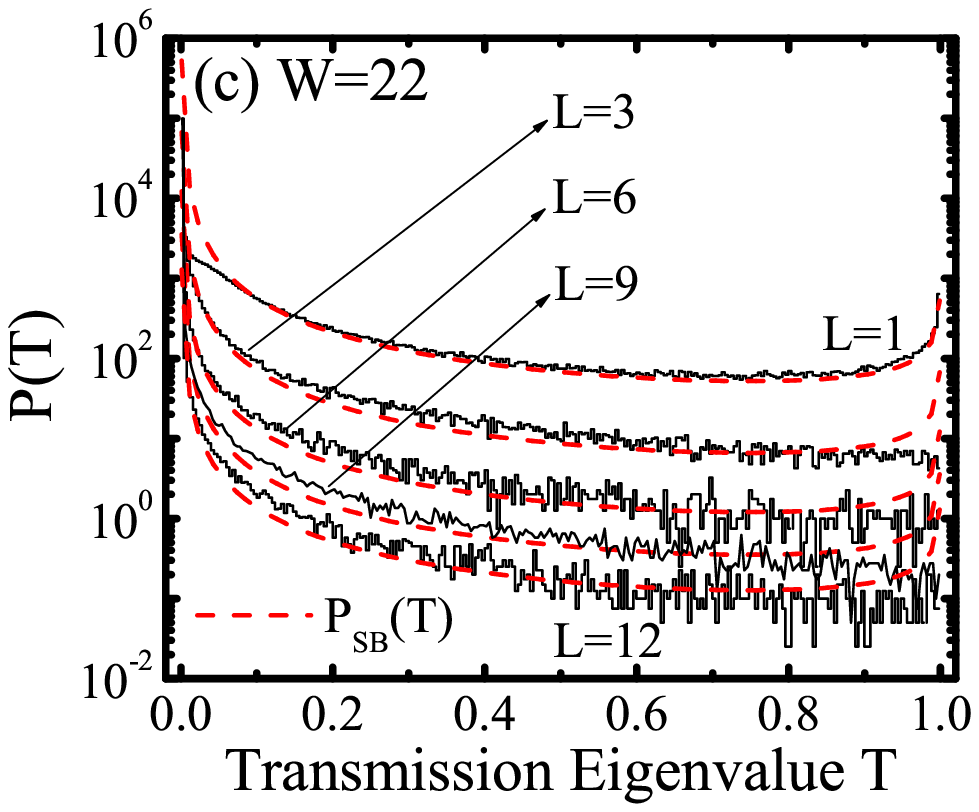,width=3.3in,height=2.5in,angle=0} }
\caption{The distribution of transmission eigenvalues $P(T)$ obtained in an
ensemble of 1000 disordered barriers, at each junction thicknesses $L$. The chosen
disorder strengths $W$ of the random potential in the Anderson Hamiltonian generate
the following systems in thick enough barriers (Fig.~\ref{fig:anderson}): (a) for $W=4$,
diffusive semiclassical metal (b) for $W=10$, bad metal (c) for $W=22$,
Anderson insulator (here we use an ensemble of 10000 barriers). The dashed and
dotted line plot the Schep-Bauer $P_{\rm SB}(T)$ and the Dorokhov $P_{\rm D}(T)$
distributions, expected to be valid for dirty interface and diffusive semiclassical
metal, respectively. Note that these are not fits, but analytical expressions
[see Eq.~(\ref{eq:dorokhov}) and Eq.~(\ref{eq:sb})] that depend on the disorder-average
barrier conductance as a single parameter.}
\label{fig:transmission}
\end{figure}

All of the results shown in Sec.~\ref{sec:trans} and Sec.~\ref{sec:shot}
are obtained by evaluating exactly the Landauer transmission matrix for
zero-temperature quantum transport in the half-filled ($E_F=0$) Anderson
Hamiltonian Eq.~(\ref{eq:tbh}) for a finite-size barrier. The disorder averaging
is performed over an ensemble containing 1000 different samples for metallic
disorder strengths $W < 16.5$ and, due to the need to search for rare events
$T_n \simeq 1$ in special configurations of disorder, for 10000 samples on the
insulating side $W \gtrsim 16.5$.
\begin{figure}
\centerline{\psfig{file=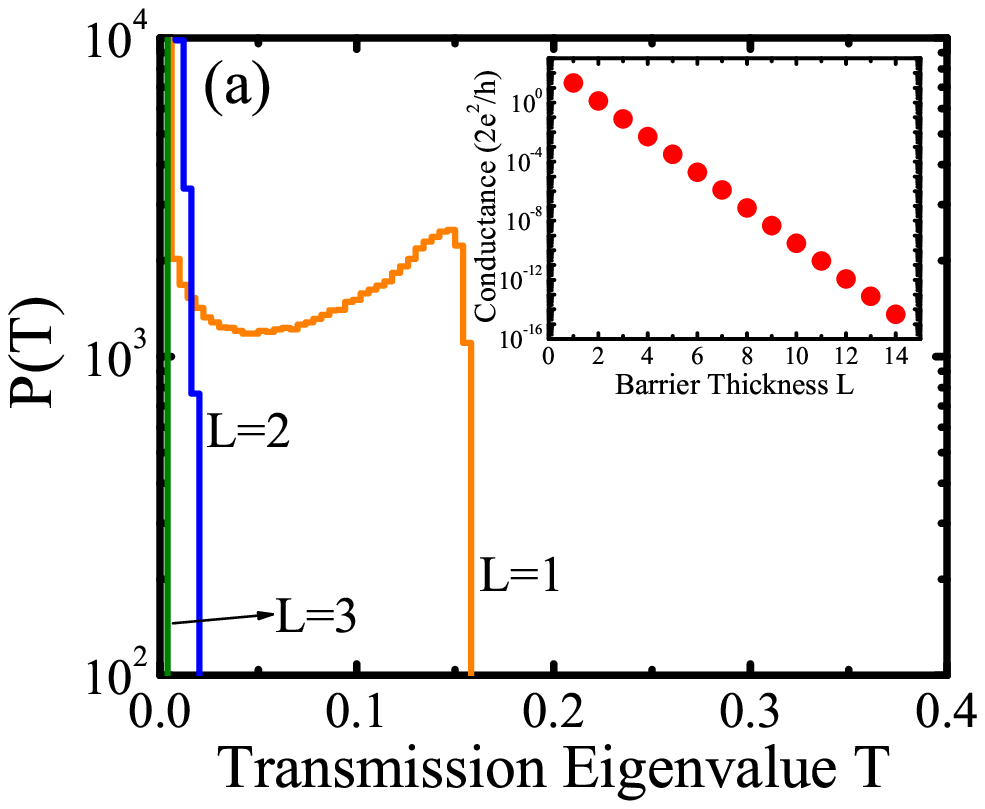,width=3.2in,height=2.2in,angle=0} }
\vspace{-0.05in}
\centerline{\psfig{file=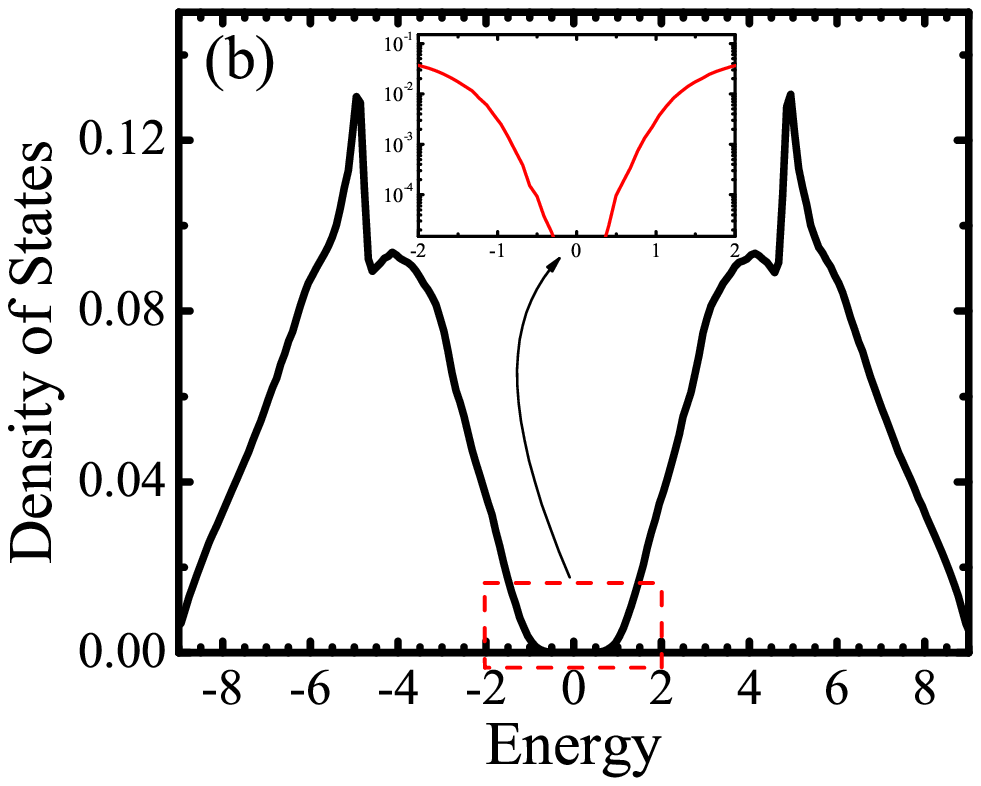,width=3.2in,height=2.2in,angle=0} }
\caption{The distribution of transmission eigenvalues $P(T)$ and
the conductance (at $E_F=0$) of tunnel barriers of different thickness that
are made of disordered binary alloy A$_{0.5}$B$_{0.5}$ [panel (a)]. The  binary
alloy is modeled by the Anderson Hamiltonian Eq.~(\ref{eq:tbh}) with
$\varepsilon_A=-\varepsilon_B=4.6$ being randomly distributed on its diagonal.
This random potential energy induces a hard gap in the eigenspectrum around $E_F=0$, as shown in the panel (b), which is phenomenologically similar to
the gap in the density of states of band or Mott insulators.} \label{fig:binary_alloy}
\end{figure}

\section{Transmission through disordered interfaces and thin barriers}\label{sec:trans}

The distribution function of the eigenvalues $T_n$ is formally defined as
\begin{equation}
P(T) = \left \langle \sum_n \delta(T-T_n) \right \rangle,
\end{equation}
where $\langle ... \rangle$ stands for averaging over all possible realizations
of impurity configurations for a given disorder strength. Early mesoscopic studies
of phase-coherent disordered conductors have been focused on bulk systems in the weak scattering regime, where one finds celebrated perturbative quantum interference effects (such as weak localization and conductance fluctuations) within diffusive transport regime.~\cite{meso} For such systems, an analytical expression for $P(T)$ has been obtained for the first time by Dorokhov~\cite{dorokhov}
\begin{equation} \label{eq:dorokhov}
P_{\rm D}(T)=\frac{\langle G \rangle }{G_Q} \frac{1} {T \sqrt{1-T}},
\end{equation}
and rederived within different theoretical frameworks.~\cite{mello,tartakovski} Here $\langle G \rangle$ is the disorder-averaged conductance. The distribution $P_{\rm D}(T)$ is  universal in the sense that it does not depend on sample-specific properties (such as dimension, geometry, and carrier-density). Although strictly derived for a quasi-one-dimensional wire (i.e., wire whose length is much bigger than its width), the scaling of transmissions implied by $P_{\rm D}(T)$ seems to have much wider validity, as long as the conductor is
in the (Bloch-Boltzmann) metallic regime.~\cite{carlo_rmt}
\begin{figure}
\centerline{\psfig{file=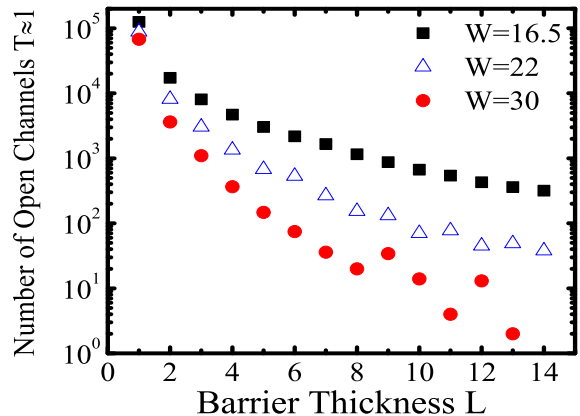,width=3.3in,height=2.3in,angle=0} }
\caption{The number of open conducting channels, whose transmission eigenvalues are
close to one $T_n \in [0.95,1]$, in an ensemble of 10000 barriers for a given junction
thickness. The barriers are made of strongly disordered materials, characterized by
the disorder strength $W =\lbrace 16.5,22,30 \rbrace$, which behaves as an Anderson
insulator in the bulk (see also related Fig.~\ref{fig:anderson} and Fig.~\ref{fig:transmission}).}\label{fig:opench}
\end{figure}

The importance of interface scattering in giant magnetoresistance phenomena~\cite{gmr_interface} has given an impetus to reexamine
transport through disordered interfaces. For the transparency of dirty
interface, whose  disorder-averaged two-probe conductance is much smaller
than the conductance of corresponding point contact $\langle G \rangle < G_{QPC}$, a Schep-Bauer distribution~\cite{schep,melsen} has been found to be applicable
\begin{equation} \label{eq:sb}
P_{\rm SB}(T)=\frac{\langle G \rangle}{\pi G_Q} \frac{1} {T^{3/2} \sqrt{1-T}}.
\end{equation}
While $P_{\rm SB}(T)$ has been derived~\cite{melsen_footnote} in the limit where barrier thickness is much smaller than the Fermi wavelength $\lambda_F$, recent experiments~\cite{naveh} have suggested that it might be valid even for thicker strongly disordered barrier $\lambda_F \ll L < \xi$, on the proviso that its width is smaller than the localization length $\xi$.

Besides diffusive wires and dirty interfaces, analytical expressions for $P(T)$ has been found for
chaotic cavities and double barrier junctions, as well as for combinations of these four generic cases.~\cite{carlo_rmt} Despite important insights obtained from different approaches~\cite{mello,carlo_rmt,dorokhov,tartakovski,schep,melsen} that yield $P_{\rm D}(T)$ and $P_{\rm SB}(T)$, no theory exist that would  make it possible to obtain explicit expression for  $P(T)$ of a 3D mesoscopic disordered conductor which is in the non-semiclassical diffusive regime (i.e., the bad metal in Fig.~\ref{fig:anderson}) extending all the way into the localized regime.~\cite{carlo_rmt}

We plot in Fig.~\ref{fig:transmission} {\em numerically exact} $P(T)$ as a function of the barrier thickness, obtained by diagonalizing ${\bf t}{\bf t}^\dag$ in Eq.~(\ref{eq:transmission}) for each sample of an ensemble of disorder configurations. For metallic diffusive barriers [panel (a)], $\ell \ll L \ll \xi$ (note that a quasi-one-dimensional system will inevitably turn into
an  insulators when $L > \xi$ independently of the strength of the disorder~\cite{mello,carlo_rmt}), transmission eigenvalue distributions follow $P_{\rm D}(T)$ prediction. However, for barriers made of the bad metal, $P(T)$ is not equal to either $P_{\rm D}(T)$ or weak-localization-corrected~\cite{nazarov} $P_{\rm D}(T)$, even though it remains bimodal distribution
with most of channels being either closed $T_n \simeq 0$ or open $T_n \simeq 1$ [panel (b)]. Note that
no single interface on the metallic side of the LD transition $W \lesssim 16.5$ can be described by
$P_{\rm SB}(T)$ [panels (a) and (b)].

When the disorder is strong enough to drive the LD transition in the bulk 3D samples, $P_{\rm SB}(T)$ becomes valid in the single plane limit [panel (c)]. Moreover, it is also useful to some extent to describe $P(T)$ for barriers composed of few such planes, as suggested by experiment and semi-intuitive arguments of  Ref.~\onlinecite{naveh}. Finally, in Sec.~\ref{sec:shot} we demonstrate that the shot noise provides very sensitive tool to compare  different distributions $P(T)$ encountered here, as well as to differentiate
those that are apparently similar [such as the distributions in panel (c)].
This is due to the fact that non-trivial features of $P(T)$, such as
the appearance of open channels in disordered tunnel
barrier,~\cite{zvi,rough} directly affect the suppression of
the shot noise power below its trivial (Poisson limit) value which
characterizes pure tunneling.

To contrast the transport through strongly disordered barriers with tunneling through barriers made of a material with a gap in the density of states (such as the conventional band-insulators determined by single-particle quantum mechanics,~\cite{rough} or more intricate Mott insulators which are governed by strongly correlated physics~\cite{freericks}), we introduce disordered binary alloy A$_{0.5}$B$_{0.5}$ between the metallic leads. This system, which is composed of an equal number of atoms $A$ and $B$ randomly distributed throughout the simple cubic lattice, is modeled by  random potential energy $\varepsilon_A = - \varepsilon_B$ on the diagonal of TBH in Eq.~(\ref{eq:tbh}). As shown in Fig.~\ref{fig:binary_alloy}, large enough $\varepsilon_A=|\varepsilon_B|$ will open a hard gap around $E_F=0$ in the density of states (DOS). The single interface of a solid with the gap in the bulk DOS can display distinctive transport properties,~\cite{freericks} which manifest here
as a non-trivial distribution $P(T)$ where non-negligible transmission eigenvalues have finite probability to appear. However, already for the ultrathin barriers $L = 4$, all transmission eigenvalues fall within the interval  $T_n \in [0,0.004]$, while the conductance exhibits typical exponential decay as a function of $L$.

On the other hand, the disordered barriers always display a non-trivial distribution of transmission eigenvalues, which can accommodate open channels even at very large $W$ and beyond the ultrathin limit. In the case of single interfaces and ultrathin barriers, the disordered region does not provide enough spatial extension in the transport direction to allow for the localization of wave functions.~\cite{schep} This  also leads to emergence of the low-energy extended states within conventional ultrathin aluminum oxide barriers that contain disorder or defects.~\cite{rough} We plot in Fig.~\ref{fig:opench} the decay of the number of open channels as the barrier thickness increases, where the disorder strengths correspond to the Anderson insulator in Fig.~\ref{fig:anderson}. The appearance of open channels beyond ultrathin barrier widths is a type of a rare event in the Anderson insulating phase (note that other types of rare events can arise in special configurations of  disorder, even in the metallic phase~\cite{rare_event}).

When Anderson insulator samples become larger than $L_\phi$, phonon-assisted
tunneling allows charges to propagate by hopping between the localized sites thereby generating a finite conductance. However, the transport studied here takes place through phase-coherent barriers (i.e., their size satisfies $L, L_y, L_z < L_\phi$)  and, therefore, effectively at zero temperature. The open channels inside the
Anderson insulator junctions are due to the tunneling via rather special configurations of localized states that provide a path for resonant transmission of electrons.~\cite{beasley} One example of such rare event is a wave function,
with energy close to the Fermi energy ($E_F=0$), which is symmetric with respect
to the leads. Such wave function would make possible resonant transmission
$T_n \simeq 1$, so that  the conductance is proportional to the probability of
finding such special barrier. This can be seen by comparing
Fig.~\ref{fig:opench} (which essentially gives the probability to encounter
an open channel in a given ensemble of barriers) to the corresponding
barrier conductances plotted in Fig.~\ref{fig:gns}.

\section{Linear statistics: Shot noise and proximity conductance} \label{sec:shot}

Over the past decade experimental and theoretical investigation of the shot
noise, as a random process characterizing non-equilibrium state into which
a phase-coherent conductor is driven by the applied voltage, has become one
of the most active frontiers in mesoscopic physics.~\cite{shot_noise} The power spectrum of the shot noise, at zero frequency and at zero temperature, can be expressed~\cite{carlo_rmt,shot_noise} in terms of the Landauer transmission
eigenvalues $T_n$ for non-interacting electrons transported through a conductor attached to
two leads
\begin{equation} \label{eq:shot_noise}
S=2 \int_{-\infty}^{+\infty} dt^\prime \, [ \overline{I(t) I(t^\prime) } - \overline{I}^2] = 2eV \frac{2e^2}{h} \sum_{n=1}^{L_y \times L_z} T_n(1-T_n).
\end{equation}
Here $\overline{I}$ is the time-average of the current flowing through the system under the
applied voltage $V$. Thus, by measuring the shot noise one effectively probes second moment
of $P(T)$, thereby obtaining complementary information to traditional conductance that is
associated with the first moment of $P(T)$.

The suppression of the shot noise power $S=2 F e \overline{I}$ with respect to the Poisson limit $S=2 e \overline{I}$ is quantified by the Fano factor $F$. In the pure tunneling regime $T_n \ll 1 \Rightarrow F=1$ because transfer of electrons through the barrier is uncorrelated in time and, therefore, described by the Poisson statistics. On the other hand, in the diffusive  metallic conductors (more precisely, in the
disordered Bloch-Boltzmann conductors in Fig.~\ref{fig:anderson} whose size is such that  $\ell \ll L$),
the shot noise power is reduced by a factor~\cite{shot_noise} $F=1/3$. This is due to the correlations
generated by Fermi statistics---electron injection into the conductor is less likely if
another electron is already occupying one of the conducting channels. In the ballistic
limit $T_n=1$, Pauli principle correlating non-interacting fermions leads to a complete noise
suppression $F=0$. The shot noise is a genuine quantum transport phenomenon since deterministic
classical transport also suppresses $S$ to zero due to the lack of stochasticity associated with
quantum mechanical propagation of electrons. The ``magic'' suppression factors, such as $F=1/3$, are expected to be universally valid, i.e., independent on the details of the system such as geometric parameters of the conductor or its resistance. Since $F=1/3$ follows from $P_{\rm D}(T)$ used in Eq.~(\ref{eq:shot_noise}), while $P_{\rm SB}(T)$ gives $F=1/2$, the Fano factors may serve as an indirect and experimentally observable confirmation of a particular distribution $P(T)$.
\begin{figure}
\centerline{\psfig{file=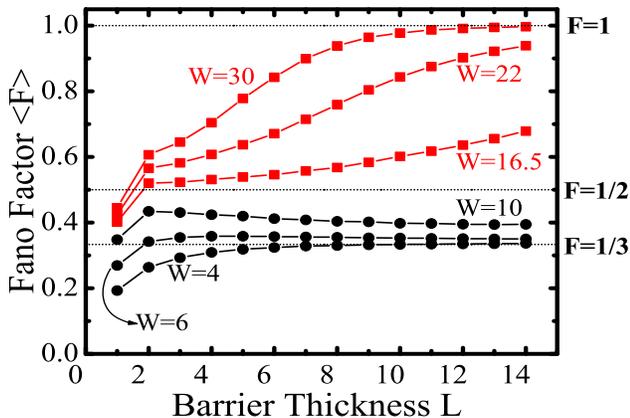,width=3.3in,height=2.3in,angle=0} }
\caption{The disorder-averaged Fano factor, quantifying suppression of
the shot noise power $S =2Fe \overline{I}$ from its Poisson value $F=1$,
as a function of the barrier thickness. Each curve is parametrized by the
strength of the disorder $W$ introduced in barrier (see Fig.~\ref{fig:anderson}).
Note that the three horizontal lines label: (i) $F=1/3$ shot noise suppression expected in the diffusive semiclassical conductors [i.e., Bloch-Boltzmann metal in Fig.~\ref{fig:anderson} whose
transparency is described by $P_{\rm D}(T)$]; (ii) $F=1/2$ for dirty interfaces
described by $P_{\rm SB}(T)$; and (iii) $F=1$ as a signature of pure tunneling through an insulator.} \label{fig:fano}
\end{figure}
\begin{figure}
\centerline{\psfig{file=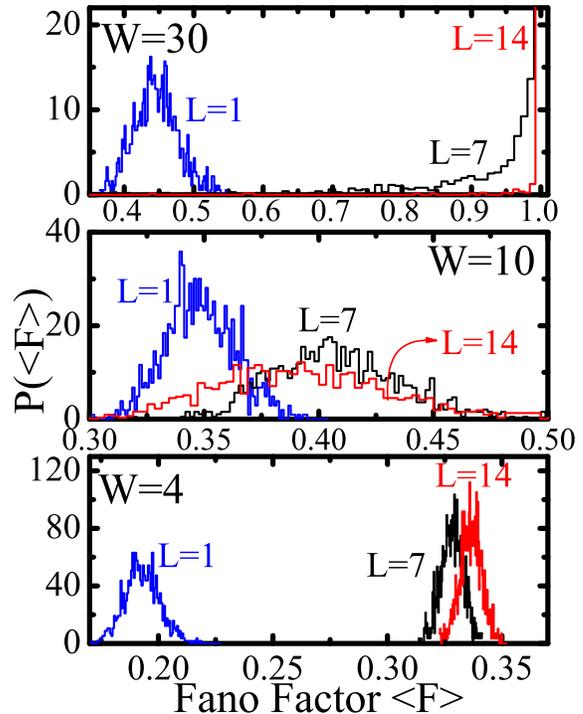,width=3.0in,angle=0} }
\caption{The full distribution function of the Fano factor for several
ensembles of barriers whose disorder-averaged $\langle F \rangle$ is plotted in  Fig.~\ref{fig:fano}.} \label{fig:fano_distribution}
\end{figure}

The knowledge of $P(T)$ makes it
possible to compute the disorder-average of any quantity that can be cast into a form of the
so-called linear statistics $A = \sum_n a(T_n)$
\begin{equation}
\left \langle A  \right \rangle = \left \langle \sum_n a(T_n) \right \rangle = \int dT \, a(T) P(T).
\end{equation}
The most frequently investigated examples of such quantities,~\cite{carlo_rmt} measured in the  two-probe geometry and at zero-temperature, are:

(a) the Landauer conductance
\begin{equation} \label{eq:landauer}
\langle G \rangle  = \frac{2e^2}{h} \left \langle \sum_n T_n \right \rangle = \frac{2e^2}{h} \int dT \, T P(T),
\end{equation}

(b) the Fano factor
\begin{equation} \label{eq:fano}
\langle F \rangle =  \frac{ \left \langle \sum_n T_n (1-T_n) \right \rangle}{\left \langle \sum_n T_n \right \rangle} = \frac{\int dT \, T(1-T) P(T)}{\int dT \, T P(T)},
\end{equation}

(c) the linear conductance of a normal-region/superconductor ($NS$) junction
\begin{eqnarray} \label{eq:gns}
\langle G_{NS} \rangle & = & \frac{2e^2}{h} \left \langle \sum_n \frac{2T_n^2}{(2-T_n)^2} \right \rangle \nonumber \\
&& = \frac{2e^2}{h} \int dT \, \frac{2T^2}{(2-T)^2} P(T),
\end{eqnarray}
which holds in the zero-voltage, zero-temperature, and zero-magnetic-field limit, and for disorder confined to the $N$ region. When the transparency of the $NS$ interface is small (e.g., due to
an insulator in between), single particle tunneling is the dominant transport mechanism which renders $\langle G_{NS}\rangle /\langle G \rangle \ll 1$ ($\langle G \rangle$ is the conductance of the
junction in the normal state). This is due to the fact that there are  no available states within
the energy gap $\Delta$ of $S$. However, in disordered-metal/superconductor junctions with transparent
$NS$ interface, $\langle G_{NS} \rangle$ is enhanced due to the proximity effect which is microscopically generated by Andreev reflection at the $NS$ interface. In  this process, an incident electron is reflected as a hole, while a Cooper pair is pushed into the superconductor. The expression for $\langle G_{NS} \rangle$ is obtained by taking into account Andreev processes via Bogoliubov-De Gennes equations, while neglecting the self-consistency issues~\cite{carlo_rmt} (e.g., superconducting order parameter is assumed to be a step function, thereby neglecting its depression on the $S$ side of the junction~\cite{nikolic_sns} as well as the terms of the order $(\Delta/E_F)^2$). For non-interacting quasiparticles that participate in purely Andreev processes at a perfectly transparent $NS$ interface $G_{NS}/G \le 2$ [the upper bound  is set by the ratio of Eq.~(\ref{eq:gns}) and Eq.~(\ref{eq:landauer}) for $T_n=1$]. Another superconducting technique which allows to experimentally probe the transparency of atomic~\cite{urbina} and mesoscopic conductors~\cite{naveh} is to sandwich them  between two superconducting leads and analyze the subharmonic gap structure of the $I-V$ characteristic of such Josephson junctions, which turns out to be determined~\cite{bardas} by $P(T)$.
\begin{figure}
\centerline{\psfig{file=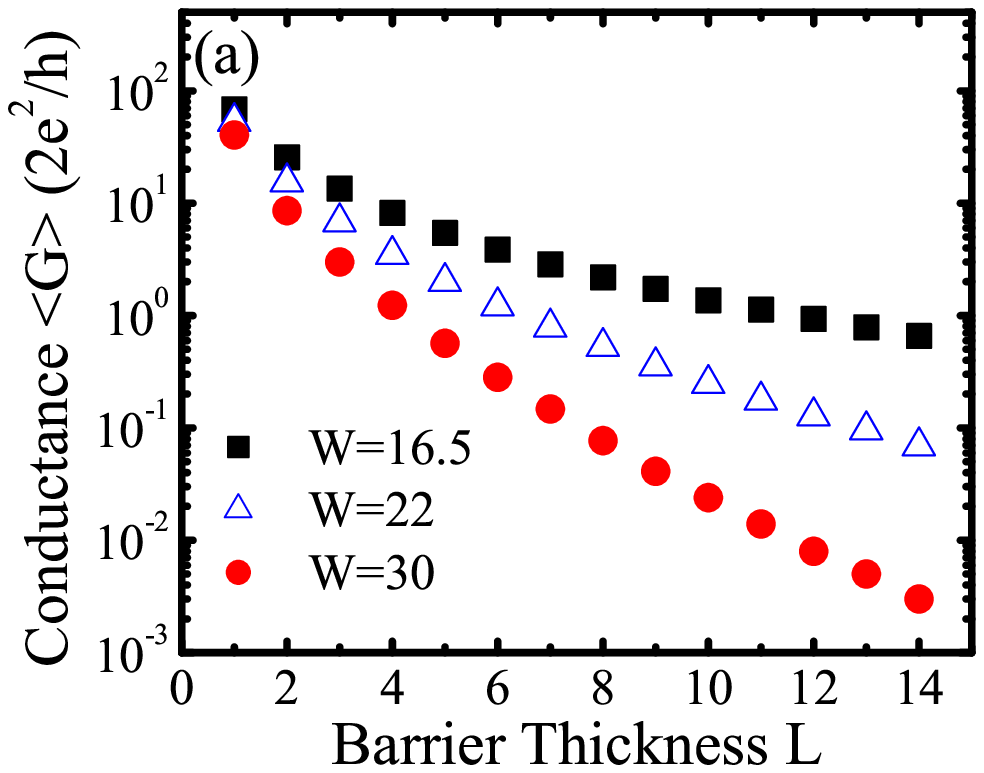,width=3.3in,height=2.3in,angle=0} }
\vspace{-0.1in}
\centerline{\psfig{file=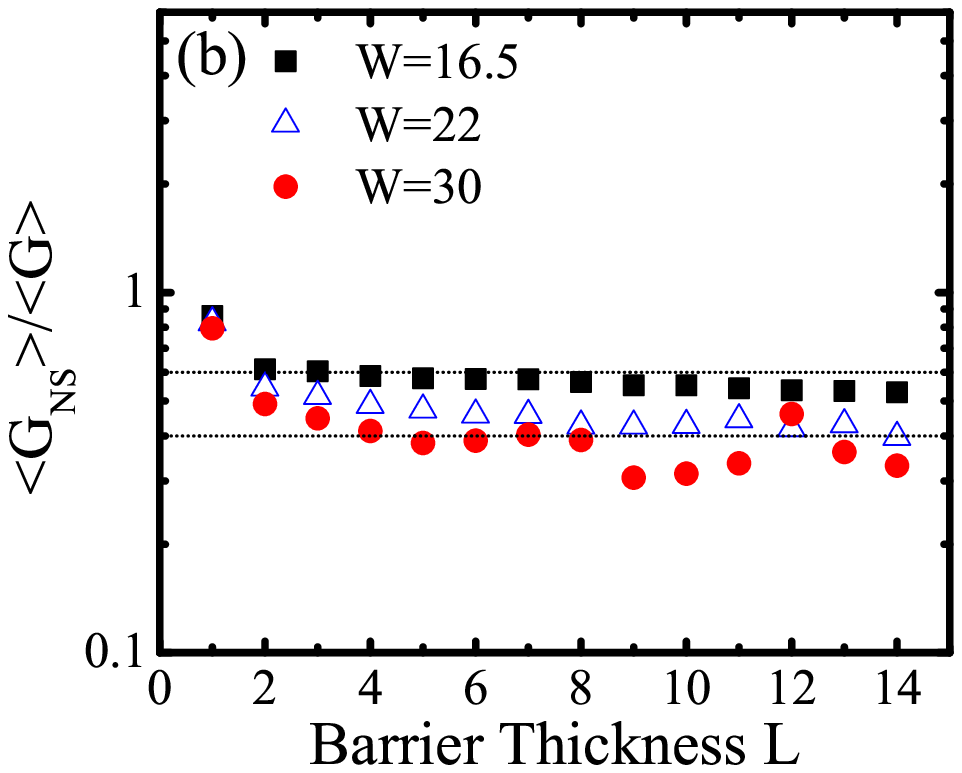,width=3.3in,height=2.3in,angle=0} }
\caption{The disorder-averaged (over 10000 samples) conductance of the Anderson insulator junctions of different thickness, attached to two metallic leads [panel (a)]. The panel (b) plots the ratio $\langle G_{NS} \rangle/\langle G \rangle$ of the disorder-averaged linear conductance of a normal-region/superconductor junction
[where normal-region is the barrier from panel (a)] and $\langle G \rangle$ from
panel (a). The dotted horizontal line serves to highlight that $\langle G_{NS} \rangle/\langle G \rangle$ is mostly confined within the interval $[0.4,0.6]$.} \label{fig:gns}
\end{figure}
\begin{figure}
\centerline{\psfig{file=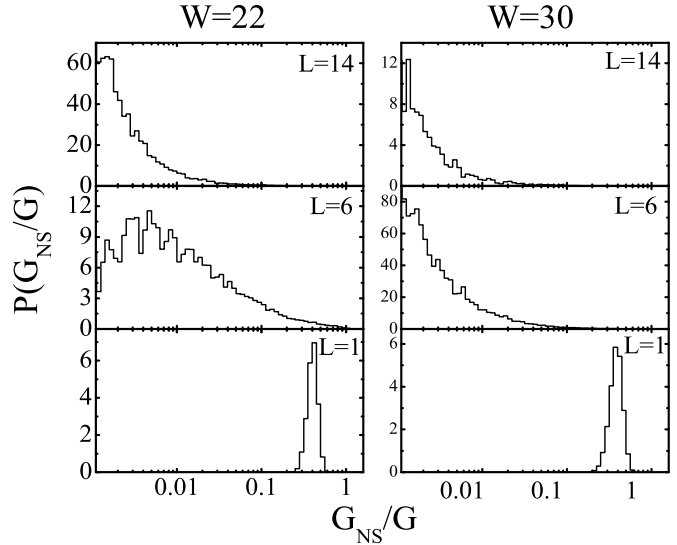,width=4.0in,height=3.0in,angle=0} }
\caption{The full distribution function (sampled over 10000 realizations of disorder) of $G_{NS} / G$ (see also Fig.~\ref{fig:gns}) quantifying mesoscopic sample-to-sample fluctuations for the Anderson insulator barriers characterized by the disorder strength $W=\lbrace 22,30 \rbrace$ and thickness $L= \lbrace 1,6,14 \rbrace$.} \label{fig:gns_mf}
\end{figure}

Figure~\ref{fig:fano} demonstrates that $F=1/3$ suppression is indeed applicable in the
Bloch-Boltzmann transport regime, i.e., in the transport through thick enough barriers (but still $L < L_y, L_z$) where semiclassical diffusive ($\ell \ll L$) charge propagation takes place. Moreover, the suppression factor saturates  $\langle F(L) \rangle \rightarrow F_M$ as a function of the barrier thickness  also for diffusion
through the bad metal barrier. However, its asymptotic value $F_M$ is steadily increasing $F_M > 1/3$ as a function of $W$ when $W \gtrsim 6$ crosses over
the boundary of semiclassical transport regime in Fig.~\ref{fig:anderson}. For
the Anderson insulator barriers $W \gtrsim 16.5$, the Fano factor is $1/3 < \langle F(L) \rangle <1$, as long as there is a probability to encounter open channels (Fig.~\ref{fig:opench}) through the barrier. It is also monotonic function of the barrier thickness since $P(T)$ scales with $L$ in a fashion shown in panel (c) of Fig.~\ref{fig:transmission}. When all channels become closed in thick Anderson insulator barriers, the Fano factor reaches its trivial asymptotic value $\langle F \rangle =1$, thereby signaling that pure tunneling takes place through such barriers. Thus, Fig.~\ref{fig:fano_distribution} suggests  that the Fano factor offers a unique single scalar quantity that is able to resolve disordered thin barriers with different transmission properties, as well as to label diffusive transport regimes of Fig.~\ref{fig:anderson} within thick barriers.

Quantum coherence, its non-local features, and randomness of microscopic details
cause large fluctuations of physical quantities in disordered mesoscopic systems.~\cite{meso} Contrary to the intuition developed from thermal fluctuations
(and their self-averaging properties) in statistical physics of macroscopic
systems, the average value and variance are not enough to characterize the distributions of various physical quantities in open (e.g., conductance, local density of states, current relaxation times, etc.) or closed (e.g., eigenfunction amplitudes, polarizability, level curvatures, etc.) mesoscopic systems. These distributions
can become particularly broad upon approaching the LD transition.~\cite{carlo_rmt} Thus, Fig.~\ref{fig:fano_distribution} introduces the full distribution function
of the Fano factor revealing that even in the diffusive metallic barriers
there are sample-to-sample fluctuations yielding wide distributions (over
the interval $F \in [0,1]$) when disorder is increased.

The possibility of fully open channels (with transparencies close to one $T_n \simeq 1$) to appear in the Anderson insulator barriers, as demonstrated in Sec.~\ref{sec:trans}, has been indirectly suggested by a zero-bias anomaly in the $I-V$ characteristic of normal-metal/Anderson-insulator/superconductor ($NIS$) junctions.~\cite{zvi} Furthermore, the comparison of the conductance $\langle G_{NS} \rangle$ of the $NIS$ junction with
the conductance $\langle G \rangle$ of the junction in the normal state  (i.e., $\langle G \rangle$
is the conductance of the metal/Anderson-insulator/metal junction) allows one to test the
importance of different special configurations of localized states that make possible
resonant tunneling responsible for the increase of the zero-bias conductance. For example, in the case of resonant tunneling through a chain of two localized sites~\cite{aleiner} $\langle G_{NS} \rangle \approx 0.27 \langle G \rangle$. The
same trend was conjectured to persist in barriers where tunneling through quasi-one-dimensional chains of arbitrary number of localized states can occur.~\cite{aleiner} As the thickness of the barrier increases, more complex configurations would allow for percolation paths
through localized states. However, they have not been observed in experiments measuring the normal conductance $\langle G \rangle$. Nevertheless, the puzzling finding of experiments~\cite{zvi} on
$NIS$ junctions is $G_{NS}/G \in [1.0,5.0]$.

Figure~\ref{fig:gns} plots the conductance of the Anderson insulator junctions, as well as the corresponding ratio $\langle G_{NS} \rangle/\langle G \rangle$ when one of the normal leads of the $MIM$ junction is turned into a superconducting one. The distributions of $P(T)$ obtained in Sec.~\ref{sec:trans} yields $\langle G_{NS} \rangle/\langle G \rangle \in [0.4,0.6]$ as a function of the barrier thickness. However, we recall here that conductance fluctuations in strongly disordered phase-coherent samples can reach the same magnitude as the conductance itself.~\cite{carlo_rmt,cf} Therefore, we investigate full distribution function of $G_{NS}/G$ in Fig.~\ref{fig:gns_mf}, which shows that particular phase-coherent samples can indeed exhibit $0.27 < G_{NS}/G < 2$ similarly to the ones found in experiments.~\cite{zvi} Nonetheless, on many $NIS$ junctions of Ref.~\onlinecite{zvi}  $G_{NS}/G > 2$  is observed. This suggests that the interplay of proximity effect in the Anderson insulator and electronic interactions~\cite{fk} (that can play an important role in the localized phase due to lack of screening) takes place. Such effects are not captured by Eq.~(\ref{eq:gns}) that takes into account only Andreev reflection of non-interacting quasiparticles at the $NS$ interface. Their treatment would require more involved theoretical approaches, such as possible combination of dynamical mean-field theory extended to inhomogeneous systems~\cite{fk} (that include  superconducting regions)~\cite{fk} with the typical medium theory of Anderson localization~\cite{vlada} which would make it possible to study proximity effect
in strongly correlated and disordered systems (modeled by the standard Hubbard
model with diagonal disorder used here).

\section{Conclusion} \label{sec:conclusion}

We have investigated how {\em statistics} of the Landauer transmission eigenvalues $P(T)$ for 3D barriers attached to two ideal metallic leads {\em scales} with the thickness of the barrier, as well as its dependence on the disorder strength which determines  different  quantum-transport regimes. When barriers are made of the bad metal (characterized by exceptionally high resistivity and lack of semiclassical mean free path), $P(T)$ remains bimodal, but it does not obey scaling predicted by the standard Dorokhov distribution. The validity of the Dorokhov distribution is
confirmed for conductors where semiclassical diffusive metallic transport takes
place, but which are not just quasi-one-dimensional wires of length much greater
than its cross section, as assumed in different theoretical derivations. The characteristic signature of the distributions of all metallic (semiclassical or quantum) diffusive barriers is encoded into the scale independent Fano factor $F \ge 1/3$  measuring suppression of the shot noise power. In special configurations of disorder, strongly disordered (i.e., Anderson insulator) barriers can accommodate fully open channels $T_n \simeq 1$ due to resonant trajectories through localized states. In experiments, this would lead to systems such as $MIM$ junctions with
$F < 1$ or normal-metal/Anderson-insulator/superconductor junctions where ratio  $G_{NS}/G$ takes any value $0 < G_{NS}/G < 2$  allowed within the proximity
theory that excludes electron correlation effects in the Anderson insulator
phase. On the other hand, the explanation of $G_{NS}/G > 2$ would require
to treat proximity effect in strongly correlated and strongly disordered systems.

\begin{acknowledgments}
We thank Z. Ovadyahu and J. K. Freericks for important insights, C. W. J. Beenakker for useful suggestion, and L. P. Z\^ arbo for help.
\end{acknowledgments}



\begin{references}
\bibitem{melsen_footnote} The analytical expression $P_{\rm SB}(T)$ was, in fact, obtained for the first time in Ref.~\onlinecite{melsen} for a system---two identical clean barriers with disorder in between---that is physically quite different form a single dirty interface.
\end{references}
\end{document}